\documentclass[12pt,preprint]{aastex}

\shorttitle{}
\shortauthors{}

\def\lya{\ifmmode {\rm Ly}\alpha~ \else Ly$\alpha$~\fi}
\def\lyan{\ifmmode {\rm Ly}\alpha \else Ly$\alpha$\fi}
\def\lyb{\ifmmode {\rm Ly}\beta~ \else Ly$\beta$~\fi}
\def\lyg{\ifmmode {\rm Ly}\gamma~ \else Ly$\gamma$~\fi}
\def\civ{\ifmmode {\rm C}\,{\sc iv}~ \else C\,{\sc iv}~\fi}
\def\civn{\ifmmode {\rm C}\,{\sc iv}~ \else C\,{\sc iv}\fi}
\def\cvi{\ifmmode {\rm C}\,{\sc vi}~ \else C\,{\sc vi}~\fi}
\def\cvin{\ifmmode {\rm C}\,{\sc vi} \else C\,{\sc vi}\fi}

\def\cv{C\,{\sc V}~}

\def\ovi{{{\rm O}\,{\sc vi}~}}

\def\ovii{{{\rm O}\,{\sc vii}~}}
\def\oviii{{{\rm O}\,{\sc viii}~}}

\def\oviin{{{\rm O}\,{\sc vii}}}
\def\oviiin{{{\rm O}\,{\sc viii}}}

\def\neixn{{{\rm Ne}\,{\sc ix}}}

\def\nexn{{{\rm Ne}\,{\sc x}}}

\def\gax{${_>\atop^{\sim}}$}

\def\chandra{{\it Chandra}~}
\def\xmm{{\it XMM-Newton}~}

\def\b{{\it b}}

\def\msun{{M$_{\odot}$}~}

\begin{document}

\title{The warm-hot gaseous halo of the Milky Way}

\author{S. Mathur\altaffilmark{1}}
\affil{Astronomy Department, The Ohio State University,
    Columbus, OH 43210}

\altaffiltext{1}{Center for Cosmology and
Astro-Particle Physics, The Ohio State University, Columbus, OH 43210}

\begin{abstract}
The circumgalactic region of the Milky Way contains a large amount of
gaseous mass in the warm-hot phase.  The presence of this warm-hot halo
observed through $z=0$ X-ray absorption lines is generally agreed upon, but
its density, path-length, and mass is a matter of debate. Here I discuss
in detail why different investigations led to different results.  The
presence of an extended (over 100 kpc) and massive (over $10^{10}$\msun)
warm-hot gaseous halo is supported by observations of other galaxies as
well.  I briefly discuss the assumption of constant density and end with
outlining future prospects.
\end{abstract}

\section{Observational evidence for the warm-hot halo}

The warm-hot gas, by definition, is the gas in the temperature range of
$10^5$--$10^7$K. At these temperatures, most of the elements are heavily
ionized and the dominant ionization states of abundant elements are
Hydrogen-like or Helium-like. The first report of the presence of such a
gas in the circumgalactic region (CGM) of the Milky Way (MW) came with
the observation of $\rm z=0$ \ovii absorption lines in the sightline
toward PKS~$2155-304$ (Nicastro et al. 2002). Since then $\rm z=0$
absorption lines in X-ray are found along several extragalactic sightlines
with one or more of \oviin, \oviiin, \neixn, \nexn, \cv or \cvi lines
(Fang et al. 2003; Rasmussen et al. 2003; McKernan et al. 2004; Williams
et al. 2005, 2006). High-resolution grating spectra on both \chandra and
\xmm facilitated the detection of these lines, but their spectral
resolution is not large enough to separate the location of the absorbing
gas into the Galactic thick disk, the CGM and/or the larger-scale local
group (LG) medium and most likely all these components contribute to the
observed column density (Mathur et al. 2008). All the observations,
however, clearly establish that {\bf the CGM of the Milky Way contains
warm-hot gas}; this result is agreed upon by most of the authors. The
discrepancy is on the extent of the warm-hot gas and its mass content.

\section{Covering fraction of the warm-hot CGM}

Several papers cited above discussed a single sightline in detail. In
order to determine the covering fraction of the warm-hot CGM,
observations of several sightlines along many different directions
through the Galactic halo are necessary and archival \chandra and \xmm
data are useful for this purpose.  Fang et al. (2006) performed a
\chandra and \xmm survey along 20 sightlines and detected $\rm z=0$
\ovii absorption in 40\% of them. Bregman et al (2007) studied 26
sightlines observed with \xmm. Gupta et al. (2012) studied 29 sightlines
with good S/N (with \ovii equivalent width (EW) limit of about 4m\AA)
and detected \ovii $\rm z=0$ absorption in 21, with resulting covering
fraction of 72\%.

\section{From equivalent width to column density}

All the studies discussed above are with absorption lines in which the
observable is the line EW. This needs to be converted to the physical
parameter of column density in order to determine the mass of the
CGM. For optically thin gas the ionic column density depends simply on
the observed equivalent width: $N(ion)=1.3\times
10^{20}(\frac{EW}{f\lambda^2})$, where $N(ion)$ is the ionic column
density ($cm^{-2}$), $EW$ is in $\AA$, $f$ is the oscillator strength of
the transition, and $\lambda$ is in $\AA$.  However, at the measured
column densities of N(\oviin), saturation could be an important issue as
suggested by simulations (Chen et al. 2003) and observational studies of
Mrk421 (Williams et al. 2005). Therefore to correctly convert the measured
equivalent widths to ionic column densities, we require knowledge of the
Doppler parameter \emph{b}; at a fixed EW, column density decreases with
increasing \emph{b}. The low velocity resolution of \chandra gratings
makes it unfeasible to directly measure the \ovii line width. If
multiple absorption lines from the same ion are detected, the relative
equivalent widths of these lines can instead be used to place limits on
the column density N(\oviin) and the Doppler parameter \emph{b} of the
medium.

Gupta et al. (2012) used this technique with \ovii K$\alpha$ and
K$\beta$ lines. For \oviin, the expected
$\frac{EW(K\beta)}{EW(K\alpha)}$ ratio is $\frac{f(K\beta)\times
\lambda^2(K\beta)}{f(K\alpha)\times \lambda^2(K\alpha)}=0.156$. Their
observations indicated that most \ovii K$\alpha$ lines are saturated, so
the inferred column densities were larger than those in optically this
case (their Table 2). 

Note that the {\bf observed EW values reported by different groups are
consistent within errors. The inferred column densities, however, are
different based on considerations for saturation}. Bregman et al. (2007)
tried to use the same technique noted above to investigate line
saturation. They missed the factor of
$\frac{\lambda^2(K\beta)}{\lambda^2(K\alpha)}$ in the equation above, so
for the expected optically thin \ovii K$\beta$ to K$\alpha$ EW ratio
they used $0.21$ instead of the correct value $0.156$. This lead them to
conclude that the lines are unsaturated, and so underestimate the column
density and the mass of the CGM. The saturation effect contributes about
a factor of four to the measured column density. The errors on \ovii
$\frac{EW(K\beta)}{EW(K\alpha)}$ ratios, however, are large and
saturation is not necessarily present in every sightline, but taking this
into account makes a difference to the average column density of the CGM.

\section{Emission measure}

The absorption lines measure the column density of gas $N_H= \mu n_e R$,
where $\mu$ is the mean molecular weight $\approx 0.8$, $n_e$ is the
electron density and $R$ is the path-length.  The emission measure, on
the other hand, is sensitive to the square of the number density of the
gas ($EM= n_e^2 R$, assuming a constant density plasma).  Therefore a
combination of absorption and emission measurements naturally provides
constraints on the density and the path-length of the absorbing/emitting
plasma. 

While the EM varies by an order of magnitude across the sky, the average
is $EM=0.0030\pm0.0006~cm^{-6}~pc$, assuming solar metallicity (Henley
et al. 2010 and Yoshino et al. 2009; see discussion on EM in Gupta et
al. 2010). Bregman et al. (2007) used $EM=0.009 cm^{-6}~pc$ for solar
metallicity which is a factor of three larger than the updated value
used by Gupta et al. This would result in a factor of three higher
density and a factor of three lower path-length compared to Gupta et
al. (2012). Fang et al. (2006) used $EM=0.0047 cm^{-6}~pc$; this would
again yield proportionately higher density and lower path-length.

\section{Density and path-length of the warm-hot CGM}

Combining the average EM$= 0.003 (\frac{Z_{\odot}}{Z}) (\frac{8.51\times
10^{-4}}{(A_O/A_H)})$ with the average absorption line column density,
we can determine the density and the path-length of the absorbing
gas. Reproducing equations 1 and 2 of Gupta et al, we have:

\begin{equation}
n_e= (2.0\pm0.6 \times 10^{-4}) (\frac{0.5}{f_{O VII}})^{-1} cm^{-3}
\end{equation}

and the path length: 

\begin{equation}
R = (71.8\pm30.2) (\frac{8.51 \times 10^{-4}}{(A_O/A_H)}) 
(\frac{0.5}{f_{O VII}})^2 (\frac{Z_{\odot}}{Z}) ~kpc
\end{equation}

where the Solar Oxygen abundance of $A_O/A_H= 8.51 \times 10^{-4}$ is
from Anders \& Grevesse (1989), $f_{O VII}$ is the ionization fraction
of \ovii and $Z$ is the metallicity.  For the observed temperature of
about \gax$10^6~$K, it is reasonable to expect $f=0.5$ (see, e.g.,
figure 4 in Mathur et al. 2003). As justified in Gupta et al,
$Z=0.3Z_{\odot}$ is a reasonable assumption. For this metallicity the
path-length becomes as large as $R=239\pm100$ kpc. As noted above, the
density is independent of metallicity.

Bregman et al. have used $A_O/A_H= 5.5 \times 10^{-4}$ and their quoted
value of the path-length, 19 kpc, is for the Solar metallicity, which is
highly unlikely in the CGM. These differences, together with column
density differences, lead to a factor of 4.6 lower path-length. Fang et
al. have used $f_{O VII}=1$ in their quoted value for the density and
solar metallicity for the quoted path-length. Rasmussen et al. (2003),
using $Z=0.3Z_{\odot}$ and $A_O/A_H= 4.6 \times 10^{-4}$ find the scale
length of the \ovii absorber to be ``at least 140 kpc'', significantly
different from the Bregman et al. and Fang et al. results.

\section{Sightline toward LMC-X3}

The nearest neighbor of our Galaxy, the Large Magellanic Cloud (LMC)
offers an unique opportunity to probe the CGM out to 50 kpc. Wang et al
(2005) present  \chandra LETG observations of LMC-X3; they detect
absorption from \ovii K$\alpha$ with EW$=20^{+14}_{-26}$ mA (90\%
confidence errors). The best-fit column density of about $10^{16}$
cm$^{-2}$ is similar to what is observed along other sightlines. Does
this suggest that the path-length of the CGM is as small as 50 kpc? Or
is the absorption from the Galactic disk as suggested by Wang et al.?

There are several reasons why this may not be the case. First of all,
the \ovii EW is not well constrained; it is consistent with zero at
$2.7\sigma$. Secondly, as noted by Wang et al., LMC-X3 is an X-ray
binary and part of the \ovii absorption may come from the outflow
arising from the binary itself. The observed \ovii column density must
have contributions from the disk, the CGM and the binary, and the total
itself can be significantly smaller than the best-fit value. Thirdly,
there is a more than factor of two uncertainty in column density
measurements toward LMC-X3, so for a constant density profile (discussed
further below) the CGM path-length may well be a over factor of two
higher, over hundred kpc.  For these reasons, the LMC-X3 sightline does
not offer additional significant insight into our understanding of the
CGM. Moreover, the LMC sightline is indeed unique, so many other
sightlines through the halo are needed to obtain the average properties
of the CGM.

\section{Galactic disk contribution to the $z=0$ absorption}

Our sightlines toward extragalactic sources pass through the Galactic
disk, but the resolution of gratings on \chandra and \xmm is not good
enough to separate out the disk and CGM components. In an effort to find
out whether most, if not all of the $z=0$ absorption arises in the
Galactic disk, Yao et al. (2008) compared an extragalactic (Mrk 421) and
a Galactic (4U $1957+11$) sightline. The 4U source is located $10$--$25$
kpc away and $2$--$4$ kpc below the plane, sampling most of the Galactic
disk in the vertical direction as well. On the other hand Mrk 421
sightline goes through the disk and the halo. Therefore the column
density difference between the two gives an estimate of the halo
contribution.

 Yao et al. find column density in the 4U direction to be
$3.1^{+5.1}_{-1.3} \times 10^{15}$ cm$^{-2}$ and in the Mrk 421
direction to be $10^{+4.7}_{-3.4}\times 10^{15}$ cm$^{-2}$. Therefore
the halo contribution is {\it at least} $10 - 3= 7 \times10^{15}$
cm$^{-2}$. This is the minimum because the column in the 4U direction is
more through plane of the disk than in the vertical direction. If the
disk is of uniform density on these scales, then the contribution from
the vertical direction is about a fifth, or $= 3/5 \times 10^{15} = 0.6
\times 10^{15}$ cm$^{-2}$. This leads to the difference of $9.4 \times
10^{15}$ cm$^{-2}$, which is the halo contribution. This is a simple,
straightforward logic and shows that most of the $z=0$ column density
toward Mrk 421 is from the halo, not from the disk.

What Yao et al. write, however, is that they have an upper limit of $4.8
\times 10^{15}$ cm$^{-2}$ for the halo contribution. How do they get
this?  They simultaneously fit the two spectra to get new values of
column densities. This gives them the column in 4U direction to be $7
\times 10^{15}$ cm$^{-2}$ with $90$\% upper limit of $12.7 \times
10^{15}$ cm$^{-2}$. Therefore they claim that the bulk of the column in
the Mrk 421 sightline is accounted for, giving the upper limit quoted
above. [Similar analysis is also done with another extragalactic source
with similar results]. Why does the joint fitting give different result
than the simple calculation mentioned above? The assumption in the joint
fit is that the gas in the disk and the halo has the same temperature
and velocity dispersion, but this assumption is unsupported.  Moreover,
their own analysis shows that this assumption is invalid; they find the
b-parameter in the 4U line to be $155$ km/s, while in the Mrk 421 line
it is $64$ km/s, therefore the two spectra should not be fit
together. Secondly, the new column toward 4U they find is {\it more} than
the $90$\% upper limit on this column from the 4U spectrum alone. This
cannot be right and is most likely the result of a much lower b-value
($= 70$ km/s) in the joint fit than in the 4U spectrum alone; for the
observed EW, a lower b-value would give higher column density. This
brings out the folly in the analysis technique of joint fitting of
unrelated spectra.


\section{Discussion}

Several authors have studied the $z=0$ absorption lines observed with
\chandra and \xmm gratings. They all agree that these lines are present
and the line EWs presented in different papers are consistent with each
other. There are, however, major differences in the final results and I
have outlined some main reasons above. They are: (1) inferred column
densities are different if saturation is not taken into account; (2)
adopted values of the average emission measure are different; (3)
adopted values of metallicity and oxygen abundance differ. Each of these
differ only by factors of few, but together make orders of magnitude
difference in the inferred mass. Since the volume goes as $R^3$, the
mass is far more sensitive to the measured path-length than the density.

Gupta et al. (2012) have taken into account line saturation, used the
most recent value for the average emission measure, adopted reasonable
values for oxygen abundance and metallicity and concluded that the CGM
in the warm-hot gaseous phase has low density (about $2 \times 10^{-4}$
cm$^{-3}$) and the path-length is over 138 kpc($1\sigma$). The inferred
electron column density is then $8.3 \times 10^{19}$ cm$^{-2}$ out to
this distance, and $=3.0\times 10^{19}$ cm$^{-2}$ out to 50 kpc. This is
well within $ 5.1 \times 10^{19}$ cm$^{-2}$ inferred from the pulsar
dispersion measure using pulsars in LMC and SMC (Taylor \& Cordes 1993).
The inferred mass of this phase of the CGM is huge, over ten billion
solar masses, comparable to the baryonic mass of the Galactic disk and
significantly more than that in any other component of the CGM. There
are, however, large uncertainties in all these estimates and they are
presented explicitly in Gupta et al. (2012). In that paper we have also
discussed all the assumptions and biases clearly and have shown that the
results are fully consistent with theoretical models.

\subsection{Assumption of constant density}

In most papers discussed above, density of the CGM is assumed to be
constant. While we have discussed this caveat in Gupta et al., it merits
additional discussion. If the density is not constant, most likely it
follows a profile such as a $\beta$-model, often observed in groups and
clusters of galaxies. In this case, the emission measure would be
sensitive to denser parts of the CGM and affect the density and
path-length estimates. In this case, the inferred path-length would in
fact be larger, and so the inferred mass. As noted above, the mass
estimate depends critically on the inferred path-length.

Secondly, the assumption of a constant density is not as bad as it may
look. In the simulations of Feldmann et al. (2012), the density is
roughly constant above the Galactic disk out to about 100 kpc. Fang et
al. (2012) also show that the hot gaseous halo of the Milky Way is
likely to have a low density extended profile as in Maller \& Bullock
(2004). Their inferred parameters are very similar to what we find in
Gupta et al. Fang et al. also note that most of the missing baryons of
the Galaxy can be in the warm-hot phase.

Thirdly, observations of other galaxies support the presence of extended
low density halo, discussed further below.

\subsection{Other galaxies}

If such a large mass of warm-hot gas exists around our Galaxy, it should
also be present around other similar galaxies. Indeed, emission from
warm-hot gas has been detected around UGC\,12591 out to 110 kpc (Dai et
al. 2011) and around NGC\,1961 out to 50 kpc. These authors calculate
halo masses out to their virial radii which are 3--6 times smaller than
the MW halo, once adjusted for the gravitational mass. First of all,
this factor is well within the uncertainties of all
measurements. Secondly, small differences in the parameters of a
beta-model can easily make a large difference when extrapolated to large
radii. Thirdly, galaxy mass many not be the relevant parameter for the
gaseous halo mass; Tumlinson et al. (2011) have shown that the specific
star formation rate of a galaxy is more important instead. Thus,
observations of other galaxies are not inconsistent with the results for
our Galaxy.

Recently Williams et al. (2012) investigated intervening X-ray
absorption line systems toward H$2356-309$ observed by Buote et
al. (2009), Fang et al. (2010) and Zappacosta et al. (2010). They found
that three of the four absorption systems originate within virial radii
of nearby galaxies or groups with projected distances of 100s of
kpc. These observations give additional evidence for extended warm-hot
halo around other galaxies. The $z=0.030$ system in Williams et al. is
particularly relevant for the present discussion because the observed
sightline passes through the halo of a nearby galaxy. The observed
column density of this absorption system is $\log
N_{OVII}=16.8^{+1.3}_{-0.9}$ at an impact parameter of D$=90$ kpc from a
nearby galaxy with virial radius of R$=160$ kpc. The path-length of the
absorber is then $2 \sqrt{R^{2}-D^{2}}=264.6$ kpc. From the path-length
and the column density we calculate the density =$7.4\times 10^{-4}
cm^{-3}$ (for \ovii ionization fraction and metallicity as in Gupta et
al. 2012). This shows that such a high density, even more than what we
calculated for the MW halo, is present out to about a hundred kpc from
another galaxy as well. This not only shows a MW-type halo around
another galaxy, it also shows that the assumption of a flat density
profile is reasonable.

\subsection{Future progress} 

 All the papers to date have used an average
emission measure for the halo; ideally we need emission measures close
to the absorption sightlines. Observations with \xmm and {\it Suzaku}
would be particularly useful in this regard. High sensitivity observations
discriminating among different halo density profiles will be a step
forward from the constant density model. Higher S/N spectra of many
sightlines will place better constraints on the column density. Thus
newer and better data and better modeling will place tighter constraints
on the physical parameters of the CGM.

There is room for progress on theory side as well. We have noted in
Gupta et al. that the observational results are consistent with recent
theoretical models. As we were about to post this article, another
theory paper on the CGM of Milky Way appeared on the arXiv by Fang,
Bullock \& Boylan-Kolchin (2012). Their results for extended hot gas
halo profiles are again consistent with Gupta et al. and they discuss
future avenues for extending their theoretical work.

It is my pleasure to acknowledge my past and present collaborators on the
topic, in particular Anjali Gupta, Yair Krongold and Fabrizio Nicastro.
  
\vspace{0.5in}
\noindent
{\bf References:}

\noindent
Anders, E. \& Grevesse, N., {\it
  Geochimica et Cosmochimica Acta}, 53, 197 \\
Bregman, J.N., \& Lloyd-Davies, E.J. 2007, \apj, 669, 990 \\
Cen, R., \& Ostriker, J. P. 1999, \apj, 514, 1 \\
Chen, X., Weinberg, D.H., Katz, N., \&  Dav\`e, R. 2003, \apj, 594, 42 \\
Fang, T.T., Canizares, C.R., \& Wolfire, M. 2006, \apj, 644, 174 \\
Fang, T.T., et al. 2010, ApJ, 714,1715 \\ 
Fang, T.T., Bullock, J., Boylan-Kolchin, M., 2012, arXiv:1211.0758 \\
Galeazzi, M., Gupta, A., Covey, K., \& Ursino, E. 2007, \apj, 658, 1081 \\
Gupta, A., Galeazzi, M., Koutroumpa, D., Smith, R., \& 
 	Lallement, R. 2009, \apj, 707, 644 \\
Hagihara, T. et al. 2010, \pasj, 62, 723 \\
Henley, D.B., Shelton, R.L.,
 	Kwak, K., Joung, M.R., \&  Mac Low, M.\-M. 2010, \apj, 723, 935 \\
Maller, A. \& Bullock, J. 2004, MNRAS, 355, 694 \\
Mathur, S., Sivakoff, G.R., Williams, R.J., \& Nicastro, F. 2008, \apss, 315, 93 \\
McCammon, D., et al. 2002, \apj, 576, 188 \\
Nicastro, F., et al. 2003, Nature, 421, 719 \\
Nicastro, F., et al. 2002, 
	ApJ, 573, 157 \\
Rasmussen, A.,
 	Kahn, S.M., \& Paerels, F. 2003, ASSL Vol. 281:
	The IGM/Galaxy Connection. The Distribution of Baryons at z=0, 109 \\
Wang, Q. D., 
	et al. 2005, \apj, 635, 386 \\
Williams, R.J.,
	 Mathur, S., Nicastro, F., Elvis, M., Drake, J.J.,
	 Fang, T., Fiore, F., Krongold, Y., Wang, Q.D., \&
	 Yao, Y. 2005, \apj, 631, 856 \\
Williams, R.J., 
	 Mathur, S., Nicastro, F., \&  Elvis, M.,2007, \apj, 665, 247 \\
Williams, R.J. et al. 2012, arXiv:1209.4080 \\
Yao, Y., Nowak, M. A., 
	Wang, Q. D., Schulz, N. S., \& Canizares, C. R. 2008, \apj, 672, 21 \\
Yao, Y., Wang, Q. D.,
 	Hagihara, T., Mitsuda, K., McCammon, D., \& 
	Yamasaki, N. Y. 2009, \apj, 690, 143 \\
Yoshino, T., Mitsuda, K.,
	 Yamasaki, N.Y., Takei, Y., Hagihara, T., Masui, K., 
  	Bauer, M., McCammon, D., Fujimoto, R., Wang, Q. D., \& 
	 Yao, Y. 2009, \pasj, 61, 805 \\

\end{document}